\documentclass{PoS}
\usepackage{calc}
\usepackage{graphicx}
\usepackage{dcolumn}
\usepackage{bm}
\usepackage{slashed}
\usepackage{amsmath,graphicx}
\usepackage{float}
\usepackage{nicefrac}
\usepackage[normalem]{ulem}
\usepackage{amsmath}
\usepackage{subfigure}
\usepackage{float} 
\usepackage{bbold}

\newcommand{\beq}{\begin{eqnarray}}
\newcommand{\eeq}{\end{eqnarray}}

\newcommand{\bel}[1]{\begin{eqnarray}\label{#1}}
\newcommand{\eel}{\end{eqnarray}}

\newcommand{\rf}[1]{Eq.~(\ref{#1})}

\newcommand{\rfn}[1]{(\ref{#1})}

\newcommand{\rff}[1]{Fig.~\ref{#1}}
\newcommand{\rfc}[1]{Ref.~\cite{#1}}

\newcommand{\p}{\partial}

\newcommand{\f}[2]{\frac{#1}{#2}}
\newcommand{\onehalf}{{\nicefrac{1}{2}}}

\renewcommand\sout{\bgroup \color{blue} \ULdepth=-.5ex \ULset}

\newcommand{\ed}{{\varepsilon}}       

\def\gmnU{g^{\mu\nu}}

\def\TmnU{T^{\mu \nu}}

\def\n0{n_{(0)}}
\def\e0{\varepsilon_{(0)}}
\def\P0{P_{(0)}}
\def\s0{s_{(0)}}

\def\umU{u^\mu}  
\def\unU{u^\nu}  
                         
\def\umL{u_\mu}

\def\unuL{u_\nu}

\def\epsLmnbg{\epsilon_{\mu\nu\beta\gamma}}

\def\epsLmnbg{\epsilon_{\mu\nu\beta\gamma}}

\def\S0iU{{\Sigma}^{0i}}

\def\g5{\gamma_5}

\def\Ot{\tilde \Omega}

\newcommand{\lp}{\left(}
\newcommand{\rp}{\right)}

\newcommand{\lsb}{\left[}
\newcommand{\rsb}{\right]}

\title{Relativistic fluid dynamics of spin-polarized systems of particles}

\ShortTitle{Relativistic fluid dynamics of spin-polarized systems of particles}

\author{Wojciech Florkowski \\
	Institute of Nuclear Physics Polish Academy of Sciences, PL-31342 Krakow, Poland,\\
	ExtreMe Matter Institute EMMI, GSI, D-64291 Darmstadt, Germany \\
E-mail: \email{Wojciech.Florkowski@ifj.edu.pl}
}
\author{Bengt Friman\\
	GSI Helmholtzzentrum f\"ur Schwerionenforschung, D-64291 Darmstadt, Germany\\
E-mail: \email{b.friman@gsi.de}
}
\author{Amaresh Jaiswal\\
	School of Physical Sciences, National Institute of Science Education and Research, HBNI, Jatni-752050, India\\
	ExtreMe Matter Institute EMMI, GSI, D-64291 Darmstadt, Germany\\
E-mail: \email{a.jaiswal@niser.ac.in}
}
\author{\speaker{Radoslaw Ryblewski} \\
	Institute of Nuclear Physics, PL-31342 Krak\'ow, Poland\\
	ExtreMe Matter Institute EMMI, GSI, D-64291 Darmstadt, Germany\\
	E-mail: \email{Radoslaw.Ryblewski@ifj.edu.pl}
}
\author{Enrico Speranza\\
	Institute for Theoretical Physics, Goethe University, D-60438 Frankfurt am Main, Germany\\
		E-mail: \email{ esperanza@th.physik.uni-frankfurt.de}
}
\date{\today}
 
\abstract{We review basic ingredients of the recently introduced perfect-fluid hydrodynamic equations for particles with spin one-half. For a quasi-realistic setup, first numerical solutions for various hydrodynamic variables including the spin polarization tensor are presented. Our results indicate that the initial spin polarization of the hottest region may play a dominant role in the polarization of particles at freeze-out.}

\FullConference{XIII Quark Confinement and the Hadron Spectrum - Confinement2018\\
		31 July - 6 August 2018\\
		Maynooth University, Ireland}

\begin{document}

\section{Introduction}
\label{sec:i}
%
The first positive measurements of global spin polarization of $\Lambda$ hyperons in heavy-ion collisions \cite{STAR:2017ckg,Adam:2018ivw} motivated intense theoretical studies of possible physical mechanisms responsible for polarization of nuclear matter created in such processes. Among others, a particularly interesting explanation of this phenomenon is based on a direct coupling between spin polarization and thermal vorticity, which was demonstrated to hold strictly in global thermal equilibrium~\cite{Becattini:2007sr,Becattini:2009wh, Becattini:2013fla, Becattini:2016gvu}. Since it is well established that the space-time behavior of matter produced in heavy-ion collisions is very well described in terms of relativistic fluid dynamics \cite{Florkowski:2010zz,Jaiswal:2016hex,Florkowski:2017olj}, it is tempting to incorporate the spin polarization evolution into a hydrodynamic scheme, which is based on the concept of local equilibrium. 

The first steps in this direction were done in Refs.~\cite{Florkowski:2017ruc, Florkowski:2017dyn,Florkowski:2018myy,Becattini:2018duy}, see also \cite{Florkowski:2017njj,Florkowski:2018ual}, where perfect-fluid hydrodynamics of particles with spin $\onehalf$ was introduced. More recent works \cite{Florkowski:2018ahw,Florkowski:2018fap} suggest an extension of the original formulation~\cite{Florkowski:2017ruc}, employing an asymmetric energy-momentum tensor and corresponding modifications of the evolution equation for the spin tensor. However, at the moment a numerical implementation is available only for the original formulation~\cite{Florkowski:2017ruc} and in this contribution the results obtained with this version of hydrodynamics with spin will be presented. We present further comments on this issue at the end of Sec.~\ref{sec:spinev}.
%
\section{Hydrodynamic background evolution}
\label{sec:hbe}
%
We first introduce evolution equations for the hydrodynamic background, which form the basis of the perfect-fluid hydrodynamics of particles with spin $\onehalf$, as formulated in~\rfc{Florkowski:2017ruc}. The conservation of energy and linear momentum requires that the divegence of energy-momentum tensor vanishes
\bel{emc}
\p_\mu \TmnU = 0\,.
\eel
One can show that for the case of spin-polarized perfect fluid the energy-momentum tensor has the structure \cite{Florkowski:2017ruc}
\bel{emt}
\TmnU = (\ed + P) \umU \unU - P \gmnU\,,
\eel
where $\ed(x)$ and $P(x)$ are the energy density and pressure, respectively, $\umU(x)$ is the fluid flow four-vector, and $\gmnU= \hbox{diag}(+1,-1,-1,-1)$ is the metric tensor. The transverse and longitudinal (with respect to $\umU$) projections of \rf{emc}  take the forms,
\beq
\dot{\ed} + (\ed + P) \theta &=& 0\,,\label{emcprojl}\\
(\ed + P)  \dot{u}^\mu &=& \p^\mu P - \umU \dot{P}\label{emcprojt}\,,
\eeq
where $\theta\equiv \p\cdot u$ is  the expansion scalar and $\dot{(\hphantom{A})}\equiv u \cdot\p $ is the comoving derivative.

Using the equilibrium distribution functions proposed in Ref.~\cite{Becattini:2013fla} with the relativistic kinetic theory expressions from Ref.~\cite{deGroot:1980}, one can express all thermodynamic quantities in terms of temperature, $T$, baryon chemical potential, $\mu$, and spin chemical potential, $\Omega$. In such a case, using the thermodynamic relation
\bel{tid}
\ed + P= s T + \Omega w +\mu n
\eel
and the expressions for  derivatives of pressure,
\bel{deriv}
s = \left.\f{\p P}{\p T}\right\vert_{\mu,\Omega}, \qquad  n = \left.\f{\p P}{\p \mu}\right\vert_{T,\Omega},\qquad  w = \left.\f{\p P}{\p \Omega}\right\vert_{T,\mu},
\eel
one can demonstrate that \rf{emcprojl} is equivalent to the equation
\bel{emcuproj}
T \p_\mu (s \umU)+\mu\, \p_\mu (n \umU)+\Omega\, \p_\mu (w \umU)=0,
\eel
where $s$, $n$, and $w$ are the entropy density, baryon number density, and spin density, respectively.  Assuming that the entropy and baryon number are conserved independently in the system, one obtains three separate conditions:
\beq
\dot{s} + s\, \theta =0 \label{entcons}\,, \qquad
\dot{n} + n\, \theta=0 \,, \qquad
\dot{w} + w\, \theta=0 \,.
\eeq

Equations~\rfn{emcprojt} and \rfn{entcons} become now six differential equations for $T$, $\mu$, $\Omega$, and three independent components of the flow $u^\mu = \gamma(1, \boldsymbol{v})$. This may be explicitly demonstrated  by using expressions for thermodynamic variables valid for a system consisting of spin $\onehalf$ particles with classical (Boltzmann) statistics, where:
\beq
\ed &=& 4 \, \cosh(\zeta) \, \cosh(\xi) \, \e0(T)\,,   \label{eneden} \qquad
P  =  4 \, \cosh(\zeta) \, \cosh(\xi) \, \P0(T)\,, \label{pres}\\
n &=& 4 \, \cosh(\zeta)  \, \sinh(\xi) \, \n0(T)\,, \label{chargden}\qquad
w  =  4 \, \sinh(\zeta)  \, \cosh(\xi) \, \n0(T)\,, \label{polden}
\eeq
and
\beq
s = 4 \, \cosh(\zeta) \, \cosh(\xi) \, \s0(T)  - 4\, \lsb \zeta  \sinh(\zeta) \, \cosh(\xi)  + \xi  \sinh(\xi) \, \cosh(\zeta) \rsb \, \n0(T). \label{entden}
\eeq
Here $\zeta = \Omega/T$, $\xi = \mu/T$, and the subscript $(0)$ denotes standard thermodynamic quantities obtained for the spin-$0$ system, see Ref.~\cite{Florkowski:2017ruc}. 

\section{Evolution of the spin polarization }
\label{sec:spinev}
%
Since the energy-momentum tensor \rfn{emt} is symmetric, the total angular momentum conservation, $\p_\alpha J^{\alpha,\beta\gamma} = T^{\beta\gamma} -  T^{\gamma\beta} + \p_\alpha S^{\alpha,\beta\gamma} = 0$, implies that the spin tensor is conserved, $\p_\lambda S^{\lambda,\mu\nu} = 0\,.$ 
%
Using the phenomenological form of the spin tensor, $S^{\lambda,\mu\nu} = \frac{w u^\lambda}{4 \zeta} \omega^{\mu\nu}$ \cite{Becattini:2009wh},  and the spin density conservation, $\dot{w} + w\, \theta=0$, one gets the evolution equations for the (rescaled) spin polarization tensor, $\bar{\omega}^{\mu\nu}=\omega^{\mu\nu}/(2\zeta)$, in the form
\bel{st}
\dot{\bar{\omega}}^{\mu\nu}   &=&0 \label{ptc}\,.  
\eel
Equation (\ref{st}) has a straightforward interpretation, namely, the components of the tensor $\bar{\omega}^{\mu\nu}$ are conserved along the fluid worldlines.

In Ref.~\cite{Florkowski:2017ruc}, the spin polarization tensor ${\bar\omega}_{\mu\nu}$ is defined in terms of the four-vectors $ {\bar k}$ and ${\bar \omega}$,
\bel{omdec}
{\bar\omega}_{\mu\nu} \equiv {\bar k}_\mu \unuL - {\bar k}_\nu \umL + \epsLmnbg u^\beta {\bar \omega}^\gamma.
\eel
where $ {\bar k}$ and ${\bar \omega}$ are required to satisfy the  conditions
\bel{ort}
{\bar k}  \cdot u = {\bar\omega} \cdot  u ={\bar\omega} \cdot {\bar k} = 0.
\eel
In addition, since $\zeta \equiv \sqrt{\frac{1}{8}{\omega}^{\mu\nu}{\omega}_{\mu\nu}}$, one also finds 
\bel{norm}
\frac{1}{2}\, {\bar\omega}_{\mu\nu}\,{\bar\omega}^{\mu\nu}= {\bar k}\cdot {\bar k} - {\bar\omega} \cdot {\bar\omega} =1.
\eel
One can check that the orthogonality and normalization conditions specified above are fulfilled, provided the initial conditions are compatible with them and Eq.~(\ref{st}) is fulfilled.

Let us note at this place that recent works \cite{Florkowski:2018ahw,Florkowski:2018fap} showed that if one wants to connect the spin tensor to the canonical one (obtained using Noether's theorem) by a pseudo-gauge transformation then one has to add an asymmetric part to the energy-momentum tensor (\ref{emc}). This part is divergence-free and does not change the form of the energy-momentum conservation, however, it changes the dynamic equation for the spin tensor. The work on such a new formulation of hydrodynamics with spin has barely begun, hence, in this contribution we restrict ourselves to the formalism presented in Ref.~\cite{Florkowski:2017ruc}.

\section{Stationary vortex solution}
\label{sec:svs}
%
In this section we consider the case of a stationary vortex with the rotation axis oriented along the $z$ direction. The hydrodynamic flow $u^\mu$ in this case is given by the four-vector~\cite{Florkowski:2017ruc} 
\bel{uvortex}
u^\mu  
= \gamma \,(1,  -  \Ot \, y,    \, \Ot \, x,  0),
\eel
with the Lorentz factor $\gamma = 1/\sqrt{1 - \Ot^2 r^2}$, where  $r = \sqrt{x^2 + y^2}$  is the distance in the transverse plane from the vortex center (without any loss of generality we assume further that angular velocity  satisfies $\Ot >0$). Due to the limited speed of light, the flow profile \rfn{uvortex} may be realized only within a cylinder with the finite radius $R < 1/\Ot$. The hydrodynamic equations are satisfied if $T$, $\mu$, and $\Omega$ are $r$-dependent and proportional to the local $\gamma$ factor~\cite{Florkowski:2017ruc}, namely
\bel{TmuOmu}
T = T_0 \gamma,  \quad \mu = \mu_0 \gamma,  \quad  \Omega = \Omega_0 \gamma,
\eel
with $T_0$, $\mu_0$, and $\Omega_0$ being arbitrary constants. 

The relevance of the solution given by Eqs.~\rfn{uvortex} and \rfn{TmuOmu}  was emphasized in Ref.~\cite{Becattini:2007sr} in the context of the global equilibrium state with rigid rotation. In this case, it was stressed that the four-vector $\beta_\mu =u^\mu/T$ is the Killing vector, i.e., it satisfies the Killing equation
\bel{Killing}
\p_\mu \beta_\nu + \p_\nu \beta_\mu = 0.
\eel 
We note that for any form of  $\beta_\mu$ one can  introduce the concept of so-called thermal vorticity \cite{Becattini:2015ska}
\bel{thermvort}
{\varpi}_{\mu\nu} \equiv -\frac{1}{2} \lp \p_\mu \beta_\nu-\p_\nu \beta_\mu \rp.
\eel
If $\beta_\mu$ satisfies \rf{Killing}, the coefficients of the thermal vorticity tensor are constants. For the stationary vortex solution described by Eqs.~\rfn{uvortex} and \rfn{TmuOmu} one may take
\bel{omBvortex}
{  \omega}_{\mu\nu}= 
\begin{bmatrix}
	0       &  0 & 0 & 0 \\
	0  &  0    & \Ot/T_0 & 0 \\
	0  & -\Ot/T_0 & 0 & 0 \\
	0  & 0 & 0 & 0
\end{bmatrix},
\eel
so that $\zeta = \Ot/(2 T_0)$, hence $\Ot = 2 \, \Omega_0$~\cite{Florkowski:2017ruc} . Moreover, in this case the only non-zero components of ${\varpi}_{\mu\nu} $ are ${\varpi}_{xy} = -{\varpi}_{yx}$. Thus, one finds that $  {\varpi}_{xy} = 2 \zeta$.

The physical realization of the stationary vortex described in this section relies on the existence of very specific boundary conditions for the fluid. For $r < R$, i.e., in the region where $T$ grows with  distance, $T \sim\gamma$, the corresponding pressure gradient plays the role of a centripetal force. However, the fluid temperature cannot grow indefinitely and for a certain value of $r$, let us say for $r=R < 1/\Ot$, some sort of external pressure, that keeps the whole system together, must be provided. In the context of heavy-ion collisions it is difficult to imagine such a source of external pressure.  Therefore, if vortices of the form discussed above exist, the flow and temperature profiles must eventually depart from the forms given by Eqs.~\rfn{uvortex} and \rfn{TmuOmu}, as the distance $r$ increases (see Ref.~\cite{Florkowski:2018ual} for the discussion of the vortex stability problem). 

\section{Numerical solutions}
\label{sec:rotdisc}

Let us now turn to a discussion of a quasi-realistic evolution of matter in the initial stages of heavy-ion collisions. Expecting that the systems formed in such collisions are hotter inside and cooler outside, we introduce an initial Gaussian distribution of the temperature 
\bel{Gaussian}
T_{\rm i} = T_0 \,\,g(x,y,z),
\eel
where $g(x,y,z)=\exp\lp -\frac{x^2}{2 \sigma_x^2}-\frac{y^2}{2 \sigma_y ^2}-\frac{z^2}{2 \sigma_z ^2}\rp$. Here we identify the beam axis with the $x$ axis, the $x\!-\!y$ plane defines the reaction plane, and the initial rotation of the  system takes place around the $z$ axis~\footnote{Note that in the standard coordinate systems used in heavy-ion collisions, the $z$-axis coincides with the beam direction, the $x\!-\!z$ plane defines the reaction plane, and created systems are expected to rotate around the $y$ axis.}. 
\begin{figure}[H] 
	\begin{center}
		\includegraphics[angle=0,width=0.9\textwidth]{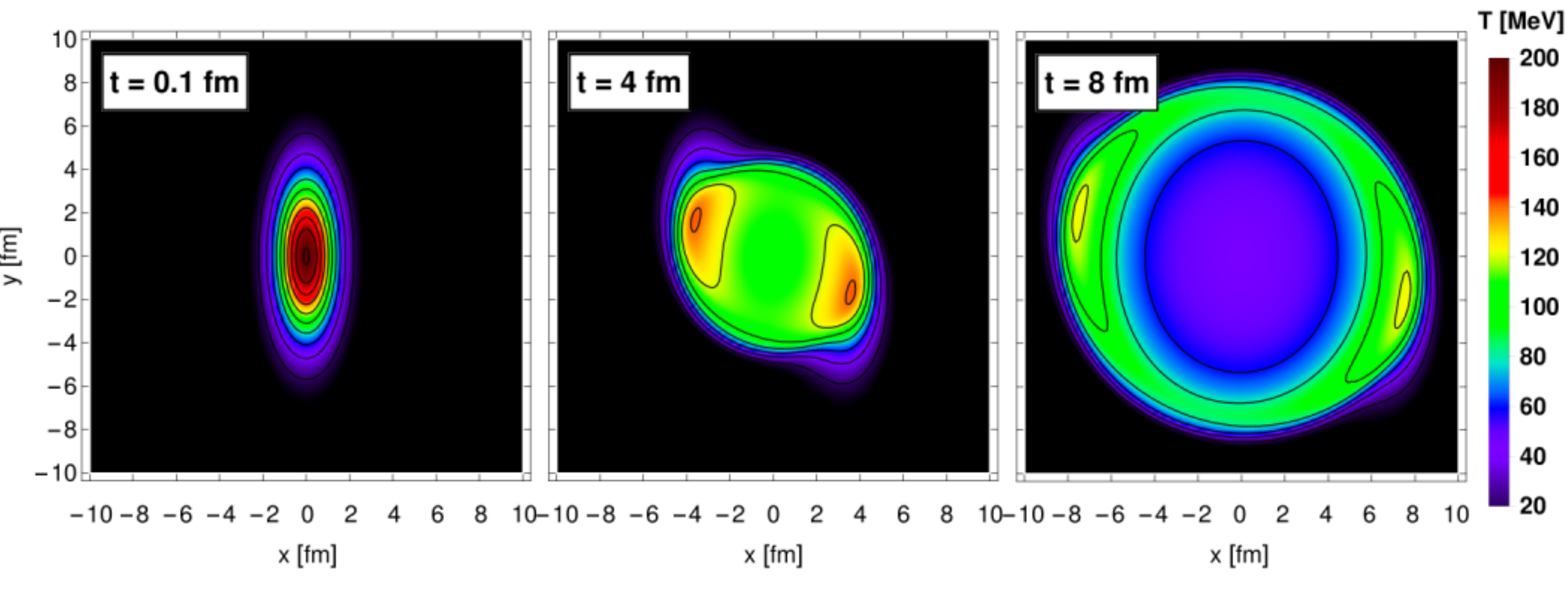}  
	\end{center} 
\vspace{-0.5cm}
	\caption{(Color online) Temperature 
		$x-y$ profiles of the system at times: $t=0.1, 4, 8$ fm (see the side bar for respective color scaling).
	}
	\label{fig:gaussfTO}
\end{figure}
The parameters $\sigma_y=2.6$~fm and $\sigma_z=2$~fm in \rf{Gaussian} are obtained from the GLISSANDO code (a numerical implementation of the Glauber model) \cite{Rybczynski:2013yba} for Au+Au collisions and the centrality class $c=20-30\%$. The value of $\sigma_x$ could be, in principle, related to the Lorentz contraction factor, however, in our calculations, for simplicity, we use the value $\sigma_x =1$~fm. The value of the initial central temperature is $T_0 = 200$ MeV, and the mass of the particles forming the fluid are chosen as the mass scale of the $\Lambda$ hyperon, $m = 1$ GeV. 

For the initial flow pattern we take the form \rfn{uvortex}, where we make the replacement $\Ot\to \lp 1/r  \rp\tanh\lp r/r_0 \rp$ with $r_0$ being extra parameter  describing the strength of the flow. In the following we use the value $r_0=1$~fm. In the limit $r_0 \to \infty$ the initial angular velocity vanishes. Following  \rfc{Becattini:2007sr} for the initial spin chemical potential we use the value $\Omega_{\rm i} = 0.03 \,T_{\rm i}/2$ so that we have $2\Omega_{\rm i}/T_{\rm i} = 0.03$. The initial baryon chemical potential profile is given by $\mu_{\rm i} = \mu_0 \,\,g(x,y,z)$, where  the  initial baryon  chemical potential at the center is $\mu_0=200$ MeV.

In Fig.~\ref{fig:gaussfTO} we show the temperature profiles of the system in the $x\!-\!y$ (reaction) plane at times: $t=0.1, 4, 8$ fm. We observe that after some time the symmetry of the initial ellipsoid elongated along the $y$ (impact vector) direction is broken. The single source splits into two separated hot spots, which move along the beam direction. In addition, the source expands and cools down. During the evolution time the entire source rotates with respect to the initial configuration.

%
\begin{figure}[H] 
\begin{center}
		\includegraphics[angle=0,width=0.9\textwidth]{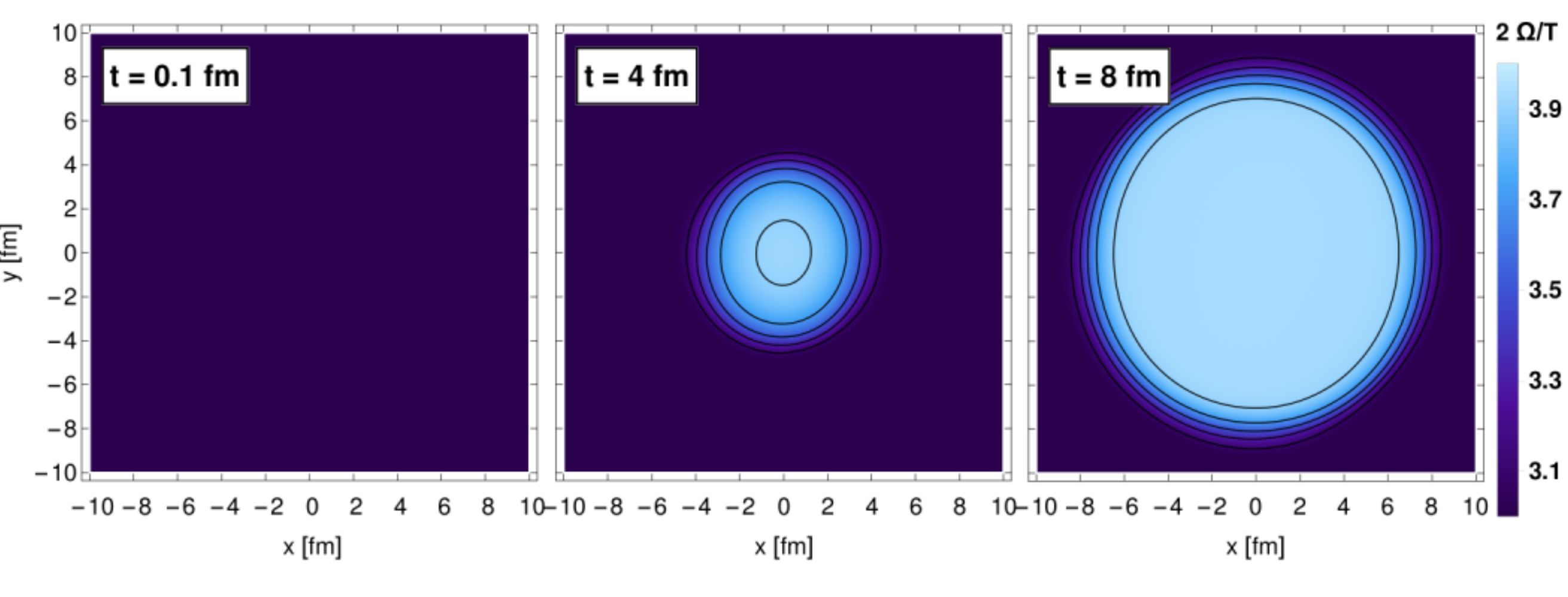} 
\end{center}    
\vspace{-0.5cm}
	\caption{(Color online) Ratio 2 $\Omega/T$ (times $100$) profiles in the $x-y$ plane of the system at times: $t=0.1, 4, 8$ fm (see the side bar for respective color scaling).  
	}
	\label{fig:gaussfZF}
\end{figure}

In  Fig.~\ref{fig:gaussfZF} we show the ratio $2\Omega/T$ multiplied by  a factor of $100$ in the $x\!-y\!$  plane at times: $t=0.1, 4, 8$ fm. Interestingly, in this case the profile is almost rotationally symmetric in the $x\!-\!y$ plane, which would suggest a stronger correlation of the $2 \Omega/T$  ratio with the initial flow profile  than with the temperature profile (see Fig.~\ref{fig:gaussfTO}).   

In the last part of this section we study the dynamics of the polarization tensor $\omega^{\mu\nu}$ on top of the hydrodynamic background evolution presented before. In order to fulfill the conditions \rfn{ort}, we assume that initially the polarization tensor is given by the expression
\bel{omB}
{\bar \omega}_{\mu\nu}= 
\begin{bmatrix}
	0       &  0 & 0 & 0 \\
	0  &  0    & {\bar \omega}_{xy} & {\bar \omega}_{xz} \\
	0  & -{\bar \omega}_{xy} & 0 & {\bar \omega}_{yz} \\
	0  & -{\bar \omega}_{xz} & -{\bar \omega}_{yz} & 0
\end{bmatrix}.
\eel
%

\begin{figure}[t!]
\begin{center}
		\includegraphics[angle=0,width=0.85 \textwidth]{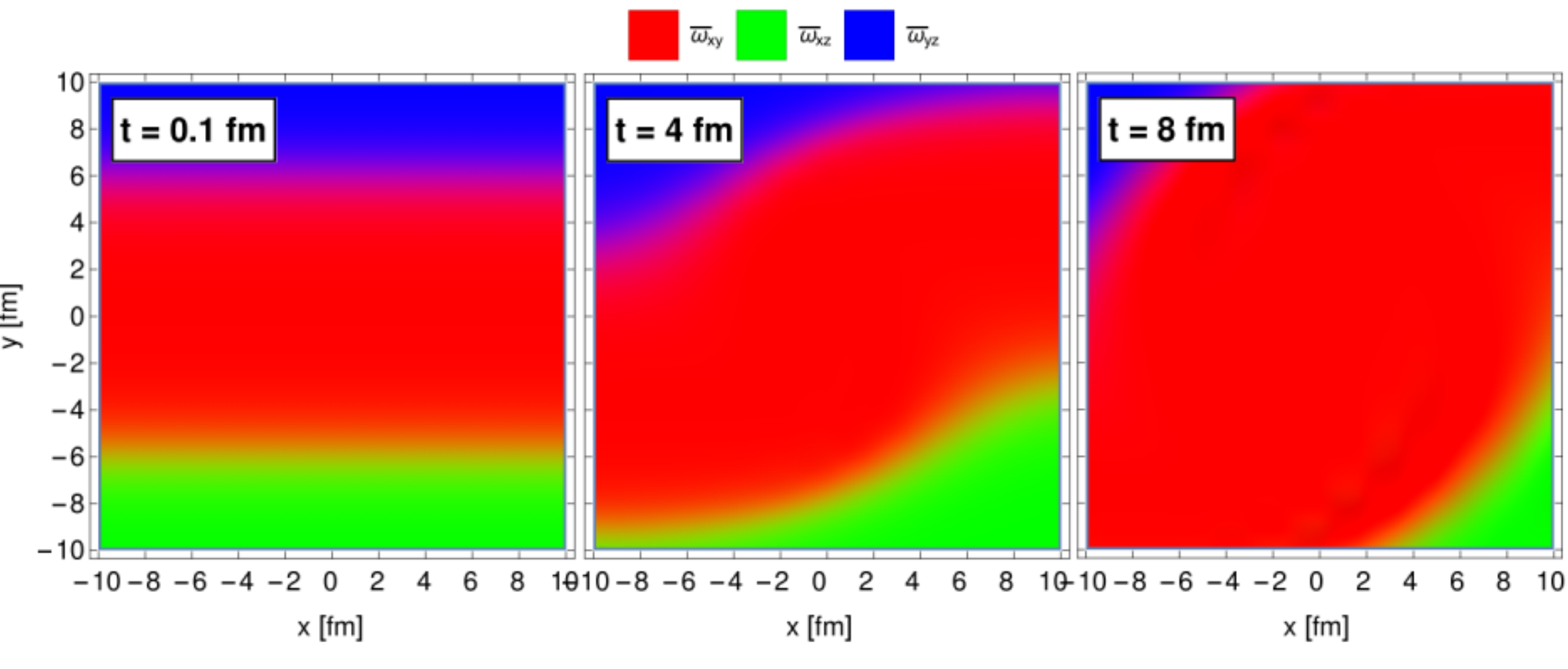}  
\end{center}
	\vspace{-0.5cm}
	\caption{(Color online)  Transverse profiles of the ${\bar \omega}_{xy}$, ${\bar \omega}_{xz}$, and ${\bar \omega}_{yz}$ components of the polarization tensor at times: $t=0.1, 4, 8$ fm (see the upper legend for respective color notation).
	}   
	\label{fig:omega}
\end{figure}
\noindent
  The normalization condition \rfn{norm} may be satisfied by adopting, for instance, the following ansatz  
\beq 
\{ {\bar \omega}_{xy},{\bar \omega}_{xz}, {\bar \omega}_{yz}\}&=& 
\begin{cases}
\{	\sqrt{(1+\tanh(y_0+y))/2}, \sqrt{(1-\tanh(y_0+y))/2},0 \}\quad y<0\\
\{	\sqrt{(1+\tanh(y_0-y))/2},0,\sqrt{(1-\tanh(y_0-y))/2} \} \quad y>0 
\end{cases}, 
\eeq
where in the numerical calculations we choose arbitrarily $y_0 = 6$. The formula above defines a simple polarization pattern in the $x\!-\!y$ plane: in the band $|y| < y_0$, the ${\bar \omega}_{xy}$ component is the dominant one, while in the region $y > y_0$ ($y < -y_0$) the ${\bar \omega}_{yz}$ (${\bar \omega}_{xz}$) component dominates.   

A subsequent evolution of these components follow from \rf{st}. In \rff{fig:omega} we present the time    evolution of the ${\bar \omega}$ components in the $x-y$ plane. One can observe that due to the fluid flow the polarized regions are transported  to different regions of space. In particular, we note that the regions rotate around the $y$ direction and expand. In this way, the ``red'' polarization region present initially in the hot centre, expands and fills the whole region with a sufficiently large temperature.  
\vspace{-0.25cm}
\section{Summary}
\label{sec:sum}

In this work we  have shown the first numerical solutions of the recently introduced perfect-fluid hydrodynamic equations for particles with spin $\onehalf$ \cite{Florkowski:2017ruc}. In this framework  the polarization direction is conserved (transported) along the stream lines. Our numerical results presented herein demonstrate this property and show that the initial  polarization of the hottest region may dominate the polarization of the particles at freeze-out.
\vspace{-0.25cm}
\acknowledgments

This work was supported in part by the ExtreMe Matter Institute EMMI at the GSI Helmholtz\-zentrum f\"ur Schwerionenforschung, Darmstadt, Germany and by the Polish National Science Center Grant No. 2016/23/B/ST2/00717. B.~F. and E.~S. were supported in part by the Deutsche Forschungsgemeinschaft (DFG) through Grant No. CRC-TR 211. A.~J. is supported in part by the DST-INSPIRE faculty award under Grant No. DST/INSPIRE/04/2017/000038. E.~S. was supported by Bundesministerium f\"ur Bildung und Forschung (BMBF) Verbundprojekt 05P2015 -- Alice at High Rate.%

\bibliographystyle{JHEP}
\bibliography{spinref}{}


\end{document}